\documentclass[journal]{IEEEtran}

\usepackage[T1]{fontenc}
\usepackage{amsmath,amssymb,amsfonts}
\usepackage{bm}
\usepackage{graphicx}
\usepackage{booktabs}
\usepackage{tikz}
\usepackage{pgfplots}
\usepackage[caption=false,font=footnotesize]{subfig}
\pgfplotsset{compat=1.17}
\usetikzlibrary{arrows.meta,positioning,calc,fit,backgrounds,shapes.misc}
\usepackage[hidelinks]{hyperref}
\usepackage{balance}
\usepackage[most]{tcolorbox}

\newcommand{\tF}{\mathrm{F}}   
\newcommand{\ti}{\mathrm{i}}   
\newcommand{\tg}{\mathrm{g}}   
\newcommand{\tr}{\mathrm{r}}   
\newcommand{\tc}{\mathrm{c}}   
\newcommand{\Jmat}{\mathcal{J}}

\newcommand{\C}{\mathbb{C}}
\newcommand{\NRMSE}{\mathrm{NRMSE}}

\tikzset{
  blk/.style={draw, rounded corners=1pt, minimum height=6.5mm, minimum width=11mm, inner sep=2pt, font=\scriptsize, fill=blue!6},
  plant/.style={draw, rounded corners=1pt, minimum height=9mm, minimum width=17mm, inner sep=2pt, font=\scriptsize, fill=violet!8},
  loopblk/.style={draw, rounded corners=2pt, minimum height=9mm, minimum width=16mm, align=center, font=\scriptsize, fill=blue!6},
  sig/.style={-{Stealth[length=1.6mm]}, semithick},
  lbl/.style={font=\scriptsize},
  slbl/.style={font=\tiny},
  sum/.style={draw, circle, inner sep=0.6pt, font=\tiny},
}

\begin{document}

\title{Recursive Self-Improvement LLM Agents for Inverter Dynamic Model Identification}

\author{Jie~Feng,
        Xiaoyang~Wang,
        Xin~Chen,
        and~Yuanyuan~Shi%
\thanks{J. Feng and Y. Shi are with the Department of Electrical and Computer Engineering, University of California San Diego, La Jolla, CA, USA (e-mail: \{jif005, yyshi\}@ucsd.edu).}%
\thanks{X. Wang and X. Chen are with the Department of Electrical and Computer Engineering, Texas A\&M University, College Station, TX, USA (e-mail: \{wangxy, xin\_chen\}@tamu.edu).}%
\thanks{This is a position paper reporting preliminary results without peer review.}}

\maketitle

\begin{abstract}
This is a position paper. We demonstrate that recursive self-improvement (RSI) large language model (LLM) agents are a natural search engine for dynamic model identification of inverter-based resources (IBRs) whose internal controls are often proprietary and hidden from the grid operators. 
White-box models provide physical transparency but require vendor disclosure; black-box models avoid this requirement but sacrifice interpretability; and existing grey-box approaches, including sparse and symbolic regression, are poorly suited to discovering feedback control architectures or incorporating control-engineering priors.
Our position is that this gap can be alleviated by (\romannumeral1) restricting the search space to a typed vocabulary of standard control modules, including PI controllers, phase-locked loops (PLLs), low-pass filters, etc., composed under block-diagram grammar rules, and (\romannumeral2) using an RSI LLM agent to perform program search over candidate block-diagram models, guided by measured frequency-domain admittance data at the point of common coupling (PCC), while fitting the free parameters of each candidate by nonlinear least squares. We instantiate this position by adapting \textsc{ThetaEvolve}, an open-source program-evolution framework supporting in-context evolution and test-time learning, to inverter model discovery. In a proof-of-concept study on a grid-following (GFL) inverter benchmark, the RSI loop reduces the normalized root mean square error (NRMSE) of a naive open-loop model from $0.470$ to $0.0435$ and identifies a 15-module closed-loop structure that closely resembles the hidden ground-truth GFL controller.
\end{abstract}


\section{Introduction}
\IEEEPARstart{P}{ower} systems are undergoing a fundamental shift toward inverter-interfaced generation. Solar photovoltaics (PV), wind turbines, battery energy storage systems, and grid-supporting devices, collectively referred to as inverter-based resources (IBRs), connect to the grid through power-electronic converters, whose dynamic behavior is largely determined by their control software. Two dominant control paradigms have emerged: \emph{grid-following} (GFL) inverters, which regulate injected current and rely on a phase-locked loop (PLL) to synchronize with an external grid, and \emph{grid-forming} (GFM) inverters, which regulate their terminal voltage and establish their own frequency reference \cite{rocabert2012control, wang2026distributed}. Recent developments have further blurred this distinction through multi-mode and hybrid GFM/GFL controllers that transition continuously between these behaviors \cite{askarian2024multimode,wang2026unified}.

The dynamic behavior of IBR-rich power systems can be reliably assessed only when system operators have access to accurate dynamic models of the connected inverters. This requirement poses a growing challenge: because inverter control designs are typically proprietary, vendors rarely disclose the internal control structures of commercial devices. As a result, operators can observe inverter behavior only through measurements at the point of common coupling (PCC), while the control mechanisms producing those measurements remain \emph{unknown} to the system operators. This creates a critical modeling gap: small-signal stability assessment, oscillation diagnosis, and interconnection studies must often rely on models that may not accurately represent the equipment deployed in the field. This gap is not hypothetical. In the 2021 Odessa disturbance, a single transformer fault in West Texas triggered the loss of more than 1{,}100\,MW of solar PV output at facilities up to 200 miles away, with PLL loss of synchronism among the leading causes; the joint North American Electric Reliability Corporation and Texas Reliability
Entity analysis further found that inverter control settings installed in the field did not always match what had been studied at interconnection \cite{nerc2021odessa}. Measurement-based model validation at the PCC is thus becoming an operational necessity, not a research convenience.

Existing approaches span white-box, black-box, and grey-box model identification. \emph{White-box} methods derive analytical impedance or state-space models from known control structures \cite{wang2018unified}. While physically transparent, they require precisely the internal information that vendors often withhold. \emph{Black-box} methods instead identify input-output behavior directly from measurements, in either the time or frequency domain. In the frequency domain, small-signal impedance or admittance responses can be obtained through terminal perturbation injection \cite{huang2009smallsignal}, and rational transfer-matrix models can be fitted using vector fitting \cite{gustavsen1999rational}. Time-domain measurements can likewise be used to identify dynamic input-output models directly from transient trajectories. More recently, machine learning and neural network approaches have been developed to handle noisy data and multiple unknown control modes \cite{huang2026learning}. Although black-box methods avoid the need for controller disclosure, the resulting pole-zero models or neural network parameters provide limited physical insight for interpretation, extrapolation, or certification.

\emph{Grey-box} identification seeks to retain physical structure while learning unknown dynamics from data. Physics-informed sparse regression, for example, can recover unknown inverter dynamics from libraries of candidate terms \cite{zheng2025pisml}, following the broader paradigm of sparse identification of nonlinear dynamics \cite{brunton2016discovering}. However, classical symbolic regression and sparse-dictionary approaches face two key limitations in inverter modeling. First, expression trees and predefined term libraries do not naturally represent the feedback interconnections that characterize inverter control systems. Second, their search mechanisms provide limited means to incorporate control-engineering knowledge, such as the strong prior that a GFL inverter contains a PLL, into the generation of candidate models. These limitations are increasingly important as grid-level stability analysis and coordinated controller design depend on accurate and interpretable inverter models \cite{ochoa2025optimal,xu2025modelfree}.

Meanwhile, a separate body of work has established large language models (LLMs) as increasingly capable tools for both power system engineering and scientific discovery. In power systems, recent studies have assessed the capabilities and limitations of LLMs in the electric energy sector \cite{majumder2024exploring}, proposed foundation-model architectures for grid data \cite{hamann2024foundation}, and developed agentic frameworks that combine foundation models with tools and structured workflows for grid analysis and operation \cite{zhang2025poweragent,wen2025xgridagent,rojas2026llms}.

In parallel, a family of systems has shown that LLMs can generate novel scientific and algorithmic artifacts when placed in iterative propose-evaluate loops that feed their own results back into later proposals. We refer to this class collectively as \emph{recursive self-improvement (RSI) LLM agents}; Section~\ref{sub:rsi} makes the structure precise. FunSearch paired an LLM with a program evaluator to discover new mathematical constructions \cite{romeraparedes2024funsearch}, while AlphaEvolve extended this paradigm to evolutionary program discovery \cite{novikov2025alphaevolve}, followed by open-source and more sample-efficient variants \cite{sharma2025openevolve,lange2025shinkaevolve}. Other systems have automated broader portions of the research process \cite{lu2024aiscientist,aygun2026expert}. More recent approaches incorporate \emph{test-time learning}, allowing the model to adapt from its own search experience on a target problem, as demonstrated by {ThetaEvolve} \cite{wang2025thetaevolve} and TTT-Discover \cite{yuksekgonul2026learning}. 
In parallel, lightweight frameworks such as {autoresearch} have popularized iterative agent--evaluation loops, alongside broader work on autonomous research workflows and harness design \cite{karpathy2026autoresearch,hogan2026alphalab}. 


Closest to our setting, LLM-SR~\cite{shojaee2025llmsr} represents equations as executable programs and uses LLM agents to guide the search, outperforming symbolic regression methods on equation discovery tasks. However, inverter identification requires a richer hypothesis class than standalone equations: inverter dynamics arise from interconnected feedback loops, internal states, and control blocks. Extending RSI LLM agents from equation programs to such \emph{feedback-interconnected control structures} remains largely unexplored and is the focus of this work.

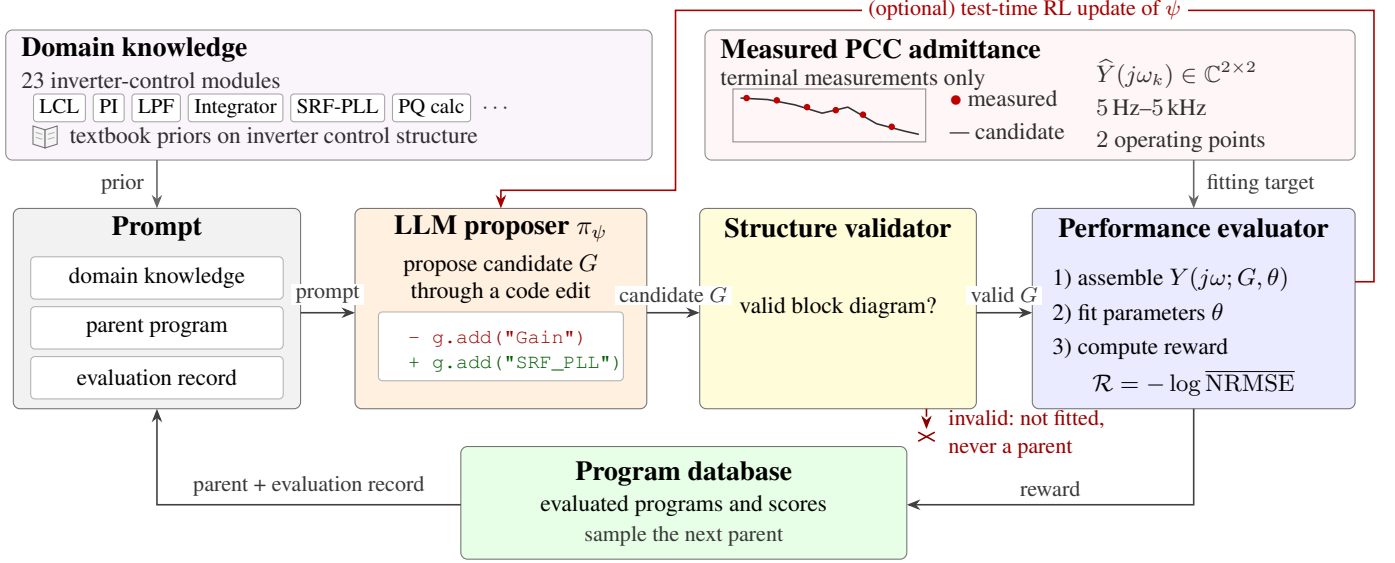
\begin{figure*}[!t]
\centering
\begin{tikzpicture}[
  x=1cm, y=1cm,
  >={Stealth[length=1.8mm,width=1.4mm]},
  pan/.style  ={draw=black!45, rounded corners=2.5pt},
  stg/.style  ={draw=black!55, rounded corners=2.5pt},
  slot/.style ={draw=black!30, rounded corners=1.6pt, fill=white,
                align=center,
                font=\fontsize{8.7}{10.2}\selectfont,
                inner xsep=3pt, inner ysep=3pt,
                minimum width=3.34cm, minimum height=5.6mm},
  ttl/.style  ={font=\fontsize{10.4}{11.8}\selectfont\bfseries, inner sep=0pt},
  bdy/.style  ={font=\fontsize{8.8}{10.3}\selectfont, inner sep=0pt, align=center},
  ann/.style  ={font=\fontsize{8.3}{9.5}\selectfont, inner sep=0pt, text=black!78},
  edgeann/.style={ann, fill=white, inner xsep=1.6pt, inner ysep=0.7pt},
  chip/.style ={draw=black!35, rounded corners=1.2pt, fill=white,
                inner xsep=2.3pt, inner ysep=1.4pt,
                text height=1.3ex, text depth=0.4ex,
                font=\fontsize{7.9}{8.9}\selectfont},
  code/.style ={font=\fontsize{7.5}{8.6}\selectfont\ttfamily},
  flow/.style ={->, semithick, black!75},
  feed/.style ={->, semithick, black!60},
  rej/.style  ={->, semithick, red!55!black, densely dashed},
  rl/.style   ={->, semithick, red!65!black},
]

\draw[pan, fill=violet!4] (0.00,5.74) rectangle (8.55,7.46);
\node[ttl, anchor=west] at (0.20,7.21) {Domain knowledge};

\node[bdy, anchor=west, text=black!82] at (0.20,6.80)
      {23 inverter-control modules};

\node[chip, anchor=west] (c1) at (0.36,6.43) {LCL};
\node[chip, right=0.9mm of c1] (c2) {PI};
\node[chip, right=0.9mm of c2] (c3) {LPF};
\node[chip, right=0.9mm of c3] (c4) {Integrator};
\node[chip, right=0.9mm of c4] (c5) {SRF-PLL};
\node[chip, right=0.9mm of c5] (c6) {PQ calc};
\node[ann, right=1.4mm of c6] {$\cdots$};

\begin{scope}[shift={(0.36,5.90)}]
  \draw[black!55, fill=black!4]
       (0,0.05) -- (0.16,0.00) -- (0.16,0.24) -- (0,0.29) -- cycle;
  \draw[black!55, fill=black!4]
       (0.16,0.00) -- (0.32,0.05) -- (0.32,0.29) -- (0.16,0.24) -- cycle;
  \draw[black!30] (0.035,0.205) -- (0.125,0.180);
  \draw[black!30] (0.195,0.180) -- (0.285,0.205);
  \draw[black!30] (0.035,0.140) -- (0.125,0.115);
  \draw[black!30] (0.195,0.115) -- (0.285,0.140);
\end{scope}

\node[bdy, anchor=west, text=black!82] at (0.84,6.04)
      {textbook priors on inverter control structure};

\draw[pan, fill=red!3] (9.25,5.74) rectangle (17.85,7.46);
\node[ttl, anchor=west] at (9.45,7.21) {Measured PCC admittance};

\node[bdy, anchor=west, text=black!82] at (9.45,6.80)
      {terminal measurements only};

\begin{scope}[shift={(9.62,5.98)}]
  \draw[black!45] (0,0) rectangle (2.55,0.70);
  \draw[semithick, black!80]
     plot coordinates {
       (0.10,0.58)
       (0.46,0.56)
       (0.82,0.49)
       (1.18,0.38)
       (1.52,0.46)
       (1.90,0.24)
       (2.28,0.14)
       (2.46,0.10)
     };
  \foreach \p in {
    (0.18,0.58),(0.58,0.55),(0.98,0.46),
    (1.36,0.42),(1.72,0.36),(2.10,0.20)
  }{
    \fill[red!75!black] \p circle (0.043);
  }
\end{scope}

\node[bdy, anchor=west, align=left, text=black!82] at (12.48,6.34)
     {\textcolor{red!75!black}{$\bullet$}\ measured\\[2pt]
      \textcolor{black!80}{\rule[0.5ex]{2.6mm}{0.5pt}}\ candidate};

\node[bdy, anchor=west, align=left, text=black!82] at (14.42,6.46)
     {$\widehat{Y}(j\omega_k)\in\C^{2\times 2}$\\[2pt]
      $5$\,Hz--$5$\,kHz\\[2pt]
      $2$ operating points};

\draw[stg, fill=black!5] (0.10,2.45) rectangle (3.90,5.10);
\node[ttl] at (2.00,4.84) {Prompt};

\node[slot] at (2.00,4.18) {domain knowledge};
\node[slot] at (2.00,3.52) {parent program};
\node[slot] at (2.00,2.86) {evaluation record};

\draw[stg, fill=orange!12] (4.62,2.45) rectangle (8.48,5.10);
\node[ttl] at (6.55,4.84) {LLM proposer $\pi_\psi$};

\node[bdy, align=center] at (6.55,4.17)
      {propose candidate $G$\\
       through a code edit};

\draw[rounded corners=1.5pt, draw=black!30, fill=white]
      (4.93,2.82) rectangle (8.17,3.73);

\node[anchor=west, code, text=red!60!black] at (5.20,3.39)
      {- g.add("Gain")};

\node[anchor=west, code, text=green!35!black] at (5.20,3.05)
      {+ g.add("SRF\_PLL")};

\draw[stg, fill=yellow!18] (9.18,2.45) rectangle (12.84,5.10);
\node[ttl] at (11.01,4.84) {Structure validator};

\node[bdy] at (11.01,3.79)
      {valid block diagram?};

\draw[stg, fill=blue!8] (13.58,2.45) rectangle (17.85,5.10);
\node[ttl] at (15.715,4.84) {Performance evaluator};

\node[bdy, align=left, anchor=west] at (13.84,4.16)
      {1) assemble $Y(j\omega;G,\theta)$};

\node[bdy, align=left, anchor=west] at (13.84,3.70)
      {2) fit parameters $\theta$};

\node[bdy, align=left, anchor=west] at (13.84,3.24)
      {3) compute reward};

\node[bdy] at (15.715,2.79)
      {$\mathcal{R}=-\log\overline{\NRMSE}$};

\draw[flow] (3.90,3.72)
      -- node[midway, above=2.8pt, edgeann]
         {prompt}
      (4.62,3.72);

\draw[flow] (8.48,3.72)
      -- node[midway, above=2.8pt, edgeann]
         {candidate $G$}
      (9.18,3.72);

\draw[flow] (12.84,3.72)
      -- node[midway, above=2.8pt, edgeann]
         {valid $G$}
      (13.58,3.72);

\draw[stg, fill=green!10] (6.02,0.46) rectangle (11.92,1.95);
\node[ttl] at (8.97,1.60) {Program database};

\node[bdy] at (8.97,1.17)
      {evaluated programs and scores};

\node[ann] at (8.97,0.77)
      {sample the next parent};

\draw[flow] (15.715,2.45)
      -- (15.715,1.18)
      -- node[midway, above=2.3pt, edgeann]
         {reward}
      (11.92,1.18);

\draw[rej] (12.18,2.45) -- (12.18,2.18);

\draw[red!55!black, semithick]
      (12.07,2.02) -- (12.29,2.14);
\draw[red!55!black, semithick]
      (12.07,2.14) -- (12.29,2.02);

\node[
  ann,
  anchor=west,
  align=left,
  text=red!60!black,
  fill=white,
  inner xsep=1pt
] at (12.43,2.12)
  {invalid: not fitted,\\[1pt]
   never a parent};

\draw[flow] (6.02,1.18)
      -- node[midway, above=2.3pt, edgeann]
         {parent + evaluation record}
      (2.00,1.18)
      -- (2.00,2.45);

\draw[feed] (2.00,5.74)
      -- node[midway, left=3pt, edgeann]
         {prior}
      (2.00,5.10);

\draw[feed] (15.715,5.74)
      -- node[midway, right=3pt, edgeann]
         {fitting target}
      (15.715,5.10);

\draw[rl] (17.85,4.13)
      -- (18.10,4.13)
      -- (18.10,7.72)
      -- (8.86,7.72)
      -- (8.86,5.34)
      -- (6.55,5.34)
      -- (6.55,5.10);

\node[
  ann,
  text=red!65!black,
  fill=white,
  inner xsep=2.4pt
] at (13.48,7.72)
  {(optional) test-time RL update of $\psi$};

\end{tikzpicture}

\caption{
The RSI loop for grey-box inverter identification.
The agent is given domain knowledge and its own previous trials; the hidden
controller is never disclosed.
At each iteration, the prompt contains the domain knowledge, a sampled parent
program, and its evaluation record.
The LLM edits the parent to propose a new candidate structure, which is first
checked for structural validity.
A valid candidate is assembled, its unknown parameters are fitted to the
measured PCC admittance, and the resulting fitting error is converted to the
reward
$\mathcal{R}=-\log\overline{\NRMSE}$,
where $\overline{\NRMSE}$ denotes the mean over the $M$ operating points.
Evaluated programs are stored in the database and subsequently sampled as
parents.
Optionally, evaluator rewards are also used for test-time RL updates of the
proposer parameters $\psi$. 
}
\label{fig:approach}
\end{figure*}

\textbf{Our position.} We argue that LLM-based reasoning is particularly well suited to identifying feedback control architectures in power systems, with inverter-model identification serving as a representative demonstration. LLMs provide a \emph{knowledge-rich search mechanism}: they can reason across physical and mathematical representations, encode substantial power and control engineering knowledge, and generate executable programs. Inverter model identification is especially well suited to these capabilities because it requires mapping physical behavior to mathematical models, using domain knowledge to search a large space of plausible control architectures, and generating executable models for direct evaluation against measurements. We therefore propose an RSI LLM agent that builds candidate inverter models by proposing and connecting standard control components into a physically meaningful block diagram, then evaluating each candidate against measured PCC admittance data. This keeps the identified model interpretable while allowing the LLM to use engineering knowledge to revise its structure, for example, adding a missing outer power-control loop when the model captures fast dynamics well but consistently fails to match the low-frequency system response.
\textbf{Contributions.} This paper makes three main contributions:
\begin{itemize}
    \item We formulate inverter model discovery as a structured search problem and develop an RSI LLM framework that iteratively proposes control structures, checks their physical validity, fits them to the PCC measurement data, and uses the resulting error to guide subsequent model refinement, as summarized in Fig.~\ref{fig:approach}. 

    \item We instantiate this framework with {ThetaEvolve} \cite{wang2025thetaevolve}, tailoring it to inverter identification with an LLM-based structural proposer drawing from a library of 23 common inverter control modules, a control structure validator, and a physics-based frequency-domain performance evaluator. We study both test-time reinforcement learning of the proposal policy and training-free search using a large off-the-shelf LLM as the agent backbone.

    \item We provide preliminary results on a simulated GFL inverter, showing that the RSI LLM agent can recover a physically meaningful closed-loop model. The identified model recovers the key synchronization, power control, and current control modules and achieves a normalized root-mean-square error (NRMSE) of $0.0435$, comparable to the $0.047$ obtained by white-box parameter fitting with the known ground-truth structure, while substantially improving over the initial NRMSE of $0.470$. 
\end{itemize}

We emphasize that this is a \emph{position paper with preliminary proof-of-concept results}, where we evaluate the proposed RSI algorithm on a simulated GFL inverter model introduced in Section~II. Our goal is to establish RSI LLM agents as a promising direction for inverter model identification, demonstrate their initial feasibility, and identify the key technical questions that remain open, as discussed in Section~\ref{sec:conclusion}. 

\section{Preliminaries and System Description}
\label{sec:prelim}
\subsection{Recursive Self-Improvement LLM Agents}
\label{sub:rsi}

By an \emph{LLM agent}, we mean a language model operating in a closed-loop with an external environment: the model proposes an action, the environment returns an observation, and subsequent actions are conditioned on the interaction history \cite{yao2023react}. Such agentic frameworks have recently been applied to power system analysis and operation, where the environment is typically a power-system tool chain \cite{zhang2025poweragent,wen2025xgridagent,rojas2026llms}. In our setting, the action is an edit to a program defining a candidate inverter model, while the environment is a deterministic evaluator that compares the candidate against PCC measurement data, and returns structured feedback and scores (e.g., fitting error) guiding subsequent model refinement.

Such a loop, by itself, is not self-improving: if every attempt is drawn
from the same fixed distribution, the agent explores but does not learn.
It becomes \emph{recursively self-improving} (RSI) when the record
generated by earlier attempts changes the distribution from which later
attempts are drawn, so that the loop's own output becomes the input that
shapes its next output. The literature has explored two main mechanisms for improving an LLM agent during search.
In the first approach, improvement is carried in the \emph{context}. The agent's previous programs, their scores, and its own written critiques are fed back into the prompt to guide subsequent proposals. This mechanism underlies self-refinement and verbal-feedback agents \cite{madaan2023selfrefine,shinn2023reflexion}, program-evolution systems that maintain populations of scored candidates \cite{romeraparedes2024funsearch,novikov2025alphaevolve}, and self-referential code-generation schemes \cite{zelikman2024stop,zhang2025darwingodel}.
In the second approach, improvement is carried in the model \emph{weights}. The same rewards used to evaluate candidate models are used to update the proposal policy during search, without requiring an external training set \cite{wang2025thetaevolve,yuksekgonul2026learning}.

In this work, we study both mechanisms for inverter model identification. Context-space recursion arises as the LLM agent evolves candidate programs using the accumulated search history, while weight-space recursion is introduced through test-time reinforcement learning that updates the proposal policy from evaluator rewards. Our current implementation uses DeepSeek-R1-0528-Qwen3-8B for weight-space adaptation; pure context-based search with a larger off-the-shelf LLM, such as Claude Opus 5, is left for future work.

\subsection{Grid-Following Inverter Model}
\label{sub:gfl}
\begin{figure}
    \centering
    \includegraphics[width=1\linewidth]{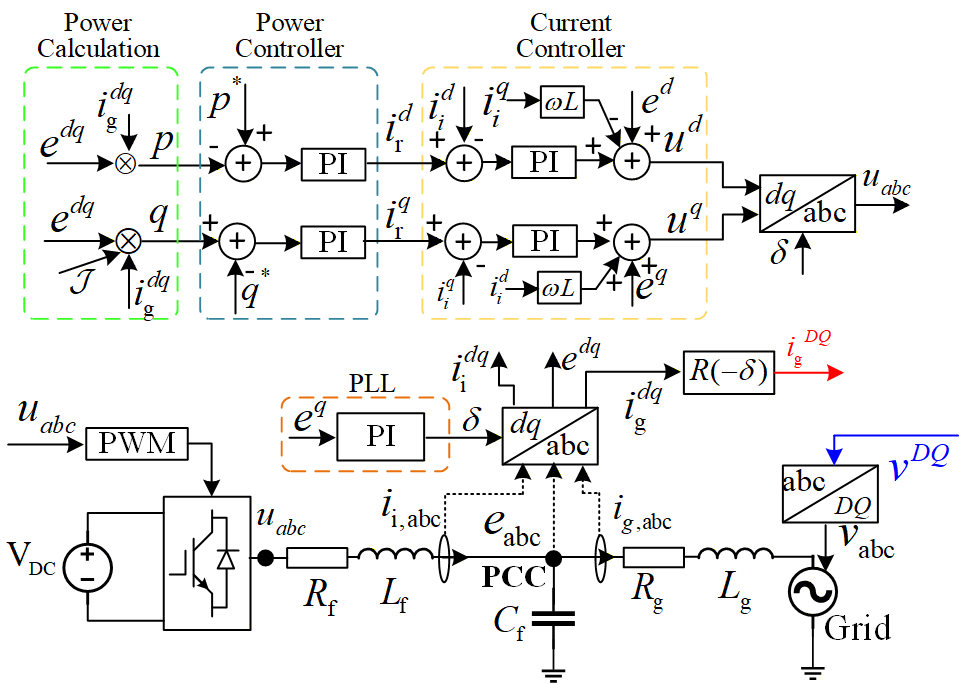}
    \caption{Canonical GFL inverter architecture considered in this
work. The terminal
small-signal relationship between $v^{DQ}$ and $i_g^{DQ}$
 is observable
at the grid.
}
    \label{fig:GFLcontrol}
\end{figure}

We ground our discussion in a standard three-phase grid-following (GFL)
inverter with an inductor-capacitor-inductor (LCL) filter, illustrated in Fig.~\ref{fig:GFLcontrol}.
The model comprises the LCL filter dynamics and three principal control
layers: a phase-locked loop (PLL) for synchronization, an outer power
controller that generates current references, and an inner current
controller that regulates the inverter current
\cite{rocabert2012control,wang2018unified}. Their interconnection forms
the layered feedback architecture that serves as a representative
structure for the identification problem considered in this work.

Throughout, $i_\ti^{dq}$ and $i_\tg^{dq}$ denote the inverter- and
grid-side currents, $e^{dq}$ the filter-capacitor voltage, $v^{dq}$ the
PCC voltage, and $u_\tr^{dq}$ the converter voltage command. Superscripts
$dq$ and $DQ$ denote quantities expressed in the inverter-local and
grid-synchronous rotating frames, respectively. We define
\begin{equation}
\Jmat =
\begin{bmatrix}
0 & -1\\
1 & 0
\end{bmatrix},
\qquad
R(\delta)=
\begin{bmatrix}
\cos\delta & \sin\delta\\
-\sin\delta & \cos\delta
\end{bmatrix},
\label{eq:rotation}
\end{equation}
where $\Jmat$ represents a $90^\circ$ rotation and $\delta$ is the
angular displacement of the inverter-local frame relative to the
grid-synchronous frame. Thus, for any two-dimensional electrical
quantity $z$,
\begin{equation}
    z^{dq}=R(\delta)z^{DQ},
    \qquad
    z^{DQ}=R(-\delta)z^{dq}.
\end{equation}

\subsubsection*{Physical layer}

The LCL filter dynamics are written compactly as 


\begin{subequations}
\label{eq:physicaldyna}
    \begin{align}
        {\dot{i}_\ti^{dq}}  & = -( \frac{\omega_0}{L_\tF}R_\tF+ {{\omega}} \mathcal{J} ) { {i}_\ti^{dq}}+ \frac{\omega_0}{L_\tF}{u_\tr^{dq}}-\frac{\omega_0}{L_\tF}{e^{dq}}  \\
       {\dot{e}^{dq}} & = - {  \omega} \mathcal{J}{  {e}^{dq}} + \frac{\omega_0}{C_\tF}{i_\ti^{dq}} -\frac{\omega_0}{C_\tF}  {i_\tg^{dq}}\\
        {\dot{i}_\tg^{dq}} & = - ({\omega} \mathcal{J} + \frac{\omega_0}{L_\tg}R_\tg) {i_\tg^{dq}} +\frac{\omega_0}{L_\tg}{e^{dq}} - \frac{\omega_0}{L_\tg}{v^{dq}} 
\end{align}
\end{subequations}
where $L_\tF, R_\tF, C_\tF, L_\tg,R_\tg$ collects the filter parameters and $\omega$
denotes the angular frequency of the inverter-local PLL frame. We assume an
LCL filter topology for the physical interface and focus on the structural
identification of the internal control architecture; the filter
parameters themselves need not be known a priori.

\subsubsection*{Synchronization}

A synchronous-reference-frame PLL synchronizes the inverter to the grid
by regulating the $q$-axis capacitor voltage $e^q$ toward zero. Let
$\phi_{\rm PLL}$ denote the PLL integrator state, $\omega$ the estimated
angular frequency, and $\omega_0$ the nominal grid frequency. A standard
PI realization is
\begin{subequations}
\label{eq:pll}
\begin{align}
    \dot{\phi}_{\rm PLL} &= e^q,\\
    \omega-\omega_0
        &= k_{\rm P,PLL}e^q
        + k_{\rm I,PLL}\phi_{\rm PLL},\\
    \dot{\delta} &= \omega-\omega_0,
\end{align}
\end{subequations}
where $k_{\rm P,PLL}$ and $k_{\rm I,PLL}$ are the proportional and
integral PLL gains. The resulting angle $\delta$ determines the frame
transformations in \eqref{eq:virotation}, while $\omega$ presents in the
physical and current-control dynamics \eqref{eq:physicaldyna}. 
\begin{equation}\label{eq:virotation}
     {{i}_\tg^{DQ}}=R(-\delta)  {{i}_\tg^{dq}}, \quad {v^{DQ}}=R(-\delta)  {v_\tg^{dq}}
\end{equation}

\subsubsection*{Outer power control}

The instantaneous active and reactive powers in per unit value are:
\begin{equation}
\label{eq:pq}
    p=(e^{dq})^\top i_\tg^{dq},
    \qquad
    q=(e^{dq})^\top \Jmat i_\tg^{dq}.
\end{equation}
Given active- and reactive-power references $p^*$ and $q^*$, the outer
PI controller maps power-tracking errors to the current reference
$i_\tr^{dq}$:
\begin{subequations}
\label{eq:pc}
\begin{align}
    \dot{\phi}^{p} &= p^*-p,
    &
    i_\tr^d
        &= k_{\rm P,PC}(p^*-p)
        + k_{\rm I,PC}\phi^p,
    \\
    \dot{\phi}^{q} &= q^*-q,
    &
    i_\tr^q
        &= -k_{\rm P,PC}(q^*-q)
        - k_{\rm I,PC}\phi^q,
\end{align}
\end{subequations}
where $\phi^p$ and $\phi^q$ are the controller integrator states and
$k_{\rm P,PC}$ and $k_{\rm I,PC}$ are the corresponding PI gains.

\subsubsection*{Inner current control}

The inner current controller tracks $i_\tr^{dq}$ using measurements of
the inverter-side current and capacitor voltage. A standard PI
realization with voltage feedforward and $dq$ cross-coupling
compensation is
\begin{subequations}
\label{eq:cc}
\begin{align}
    \dot{\gamma}^{dq}
        &=
        \omega_0
        \big(i_\tr^{dq}-i_\ti^{dq}\big),
        \\
    u^{dq}
        &=
        k_{\rm P,i}
        \big(i_\tr^{dq}-i_\ti^{dq}\big)
        + k_{\rm I,i}\gamma^{dq}
        + e^{dq}
        + \frac{\omega}{\omega_0}
          L_\tF\Jmat i_\ti^{dq},
\end{align}
\end{subequations}
where $\gamma^{dq}$ is the current-controller integrator state,
$k_{\rm P,i}$ and $k_{\rm I,i}$ are the PI gains, and $L_\tF$ is the
inverter-side filter inductance.

Together, the physical and control layers define a structured nonlinear
dynamical system
\begin{subequations}
\label{eq:nonlinear_ibr}
\begin{align}
     \dot{x}
    &=
    f(x,v^{DQ},r;\theta),
    \qquad
    r=
    \begin{bmatrix}
        p^* & q^*
    \end{bmatrix}^{\!\top},\\
    i_\tg^{DQ} &=g(x).
\end{align}
   
\end{subequations}
where $x$ collects the physical states and the internal states of the
PLL, power controller, and current controller, while $\theta$ collects
the corresponding physical and control parameters. Importantly,
$f$ is generated by an interconnection of recognizable physical and
control modules rather than by an arbitrary nonlinear mapping. $g(x)$ is the observe equation given by  \eqref{eq:virotation}.

\subsubsection*{Small-signal PCC admittance}

Although the internal controller states and interconnections are
generally unavailable to the system operator, their aggregate effect is
observable through the inverter's terminal response. Consider an
equilibrium corresponding to a fixed operating point $r=[p^*,q^*]^\top$.
Linearizing \eqref{eq:nonlinear_ibr} and the corresponding output
equation in the grid-synchronous $DQ$ frame gives  
\begin{subequations}
    \begin{align}
    \Delta\dot{x}
        &= A\Delta x+B\Delta v^{DQ},
        \\
    \Delta i_\tg^{DQ}
        &= C\Delta x.
\end{align}
\end{subequations}

The resulting frequency-domain transfer matrix from the input $\Delta v^{DQ}$ to the output $\Delta i_\tg^{DQ}$ is:
\begin{subequations}
    \begin{align}
        \Delta i_\tg^{DQ}(s)
    &=
    Y(s)\Delta v^{DQ}(s),\\
    Y(s)
    &=
    C(sI-A)^{-1}B
    \in\mathbb{C}^{2\times2}.\label{eq:admittance}
    \end{align}
\end{subequations}
This is later referred to as admittance because its input and output are voltage and current, respectively, analogous to the conventional notion of electrical admittance.

The frequency response $Y(j\omega)$ can be obtained experimentally by
injecting small voltage perturbations at the grid side 
resulting current response
\cite{wang2018unified,huang2009smallsignal}. It therefore provides an
externally measurable signature of the hidden inverter
dynamics. Synchronization and outer-control dynamics predominantly
shape the lower-frequency response, while the inner current controller
and filter dynamics become increasingly important at higher frequencies.
This paper considers the identification problem of recovering the feedback-interconnected control structure that reproduces the measured terminal admittance. 

\section{Problem Statement}
\label{sec:problem}

We consider the problem of identifying the internal feedback control architecture of an inverter from measurements available only at its grid interface. The challenge is that many different internal control structures can produce similar terminal behavior, while the controller states, signals, and interconnections are not directly observable. Our goal is therefore to recover an interpretable small-signal control model that is consistent with measured PCC admittance data. 

At $M$ known operating points, let 
$r_m=[p_m^*,q_m^*]^\top$ denote the active and reactive power setpoints of the \(m\)-th operating point. At each $r_m$, the operator measures the small-signal
admittance \eqref{eq:admittance} over frequencies
$\{\omega_k\}_{k=1}^{K}$, obtaining
\begin{equation}
    \widehat{Y}_m(j\omega_k)\in\C^{2\times2}\,, 
    k=1,\ldots,K .
\end{equation}
In this work, measurements span $5\,\mathrm{Hz}$--$5\,\mathrm{kHz}$ with $K=16$.

Let $\mathcal{M}$ denote a library of standard control components, introduced in Section~IV, including PI controllers, filters, PLLs, and power-control blocks, and let $\mathcal{G}$ denote the set of all valid control structures formed by selecting components from $\mathcal{M}$ and interconnecting them according to the prescribed block-diagram rules. A candidate inverter model is represented by a structure $G\in\mathcal{G}$.

For a given structure $G$, let $\theta_m\in\Theta(G)$ denote its continuous model parameters at operating point $r_m$, where $\Theta(G)$ denotes the feasible parameter set determined by the parameter ranges of the modules in $G$. Given $G$ and $\theta_m$, the predicted PCC admittance is
\begin{equation}
Y_m(j\omega_k;G,\theta_m)\in\mathbb{C}^{2\times2},
\qquad k=1,\ldots,K.
\end{equation}

We then define the discrepancy
between the predicted and measured admittances using the normalized
root-mean-square error (NRMSE), defined as
\begin{equation}
\label{eq:nrmse}
\NRMSE_m(G,\theta_m)
=
\sqrt{
\frac{
\displaystyle
\sum_{k=1}^{K}
\left\|
Y_m(j\omega_k;G,\theta_m)
-
\widehat{Y}_m(j\omega_k)
\right\|_F^2
}{
\displaystyle
\sum_{k=1}^{K}
\left\|
\widehat{Y}_m(j\omega_k)
\right\|_F^2
}
}.
\end{equation}
Here, $\|\cdot\|_F$ denotes the matrix Frobenius norm.

The objective is to identify a single control structure $G$ that explains the measured PCC admittance across all operating points. For each candidate structure, the continuous
parameters are fitted separately at each operating point by nonlinear
least squares. This gives the bilevel identification problem
\begin{subequations}
\label{eq:problem}
\begin{align}
\min_{G\in\mathcal{G}}\quad
&\frac{1}{M}\sum_{m=1}^{M}
\NRMSE_m\left(G,\theta_m^\star(G)\right),
\label{eq:problem-out}
\\
\mathrm{s.t.}\quad
&\theta_m^\star(G)=
\arg\min_{\theta_m\in\Theta(G)}
\NRMSE_m\left(G,\theta_m\right),\nonumber\\
&\quad m=1,\ldots,M,
\label{eq:problem-inner}
\\
&\underline{n}
\le n(G)
\le \overline{n}.
\label{eq:problem-n}
\end{align}
\end{subequations}
Here, $n(G)$ denotes the dynamic order
of the candidate model, and $\underline{n}$ and $\overline{n}$ are
prescribed complexity bounds. 
The structure $G$ is shared across all operating points, while the fitted
parameters $\theta_m^\star(G)$ may vary with the operating condition
$r_m$. At each operating point, a single parameter vector
$\theta_m^\star(G)$ is fitted jointly over all sampled frequencies
$\{\omega_k\}_{k=1}^{K}$.

The RSI LLM agent addresses the outer structural search
in~\eqref{eq:problem-out} by proposing and refining candidate structures
$G$. For each proposed structure, a parameter-identification subroutine
solves the inner nonlinear least-squares problem in~\eqref{eq:problem-inner}
to obtain $\theta_m^\star(G)$ at each operating point. The average of the resulting fitting error is used to score the candidate and construct the reward returned to the search algorithm. A candidate is favored only if it can reproduce the measured admittance over all sampled frequencies $\{\omega_k\}_{k=1}^K$ and across all operating points $\{r_m\}_{m=1}^M$. The solution model provides a shared structural description of the measured inverter, while its continuous parameters remain operating-point specific and must be refitted when applied to a new operating point.

\section{Inverter Model Search Space and Evaluation}
\label{sec:model_search}

The bilevel formulation in Section~\ref{sec:problem} separates inverter
identification into a discrete structural search over
$G\in\mathcal{G}$ and continuous parameter fitting for each candidate
structure. This section defines the structured search space
$\mathcal{G}$ and the evaluator used to solve the inner problem
\eqref{eq:problem-inner}. The recursive search over $\mathcal{G}$ is described in
Section~\ref{sec:rsi_search}.

\subsection{Module Vocabulary and Interconnection Grammar}
\label{sub:vocab}

The module library $\mathcal{M}$ consists of the $23$ module kinds
listed in Table~\ref{tab:vocab}. Each module kind specifies its input
and output ports and associated signal types, its free parameters and
their admissible ranges, its contribution to model order, and its
small-signal frequency-domain realization. For example, the transfer function of PI and LPF are defined as follows,
\begin{equation}
    M_{\rm PI}(s)
    =
    k_{\rm P}+\frac{k_{\rm I}}{s},
    \qquad
    M_{\rm LPF}(s)
    =
    \frac{\omega_\tc}{s+\omega_\tc},
\end{equation}
where $k_{\rm P}$ and $k_{\rm I}$ are PI gains and $\omega_\tc$ is the
cutoff frequency. The SRF-PLL module implements the linearization of
\eqref{eq:pll}, while the rotation modules implement
\eqref{eq:rotation} linearized about the fitted equilibrium angle. Ports
use three signal types: scalar, $dq$ vector, and power pair. Complete definitions of all modules are provided in Appendix \ref{app:vocab}.

\begin{table}[!t]
\caption{Module Vocabulary $\mathcal{M}$}
\label{tab:vocab}
\centering
\scriptsize
\setlength{\tabcolsep}{3pt}
\begin{tabular}{@{}p{0.22\columnwidth}p{0.60\columnwidth}c@{}}
\toprule
Class & Module kinds & Order \\
\midrule
Physical plant &
\texttt{LCL}, \texttt{LFilter} & $6$, $2$ \\
Dynamic control &
\texttt{PI}, \texttt{PI\_dq}, \texttt{LPF1}, \texttt{LPF1\_dq},
\texttt{Integrator} & $1$--$2$ \\
Synchronization &
\texttt{SRF\_PLL} & $2$ \\
Operating point &
\texttt{Rot\_v}, \texttt{Rot\_i}, \texttt{PQ\_calc} & $0$ \\
Static algebra &
\texttt{Gain}, \texttt{Gain\_dq}, \texttt{Negate}, \texttt{Negate\_dq},
\texttt{PackDQ}, \texttt{SliceD}, \texttt{SliceQ}, \texttt{JMul},
\texttt{ConstZero} & $0$ \\
Residual &
\texttt{Pade1}, \texttt{Pade1\_dq}, \texttt{Notch} & $1$--$2$ \\
\bottomrule
\end{tabular}
\end{table}

The admissible set $\mathcal{G}$ is defined by the module library $\mathcal{M}$ together with an interconnection grammar. Candidate structures are formed by connecting module outputs to module inputs. For example, the PCC voltage $v^{DQ}$ can be connected to the input of \texttt{Rot\_v}, which transforms the voltage from the global $DQ$ frame to the local $dq$ frame; its signal type remains a $dq$ vector. Connections are allowed only between compatible signal types, so a $dq$-vector output can only be connected to an input that accepts a $dq$ vector. Each input port is driven by exactly one source. When several signals need to be combined, the combination is represented explicitly and realized as a summation during model assembly. 

Additional structural requirements are imposed at the graph level. Each candidate must define the terminal-current output and include a physical filter module that represents the interface between the inverter and the AC grid. The dynamic order must also satisfy the bound in \eqref{eq:problem-n}.

\subsection{Frequency-Domain Evaluation and Structured Feedback}
\label{sub:eval}

For each grammar-valid candidate structure $G\in\mathcal{G}$, the evaluator assembles the linear equations defined by the module transfer functions and their interconnections. At operating point $r_m$ and frequency $s=j\omega_k$, the resulting model yields the predicted PCC admittance
\begin{equation}
Y_m(j\omega_k;G,\theta_m)
\in\C^{2\times2},
\end{equation}
where $\theta_m$ denotes the continuous model parameters to be fitted at operating point $r_m$.

\subsubsection{Parameter Fitting}

For each operating point $r_m$, the evaluator solves the inner
regression problem in \eqref{eq:problem-inner},
\begin{equation}
\label{eq:fit}
    \theta_m^\star(G)
    =
    \arg\min_{\theta_m\in\Theta(G)}
    \NRMSE_m(G,\theta_m).
\end{equation}
The structure $G$ is fixed during this optimization, and $\theta_m^\star(G)$ is fitted jointly over all frequencies $\{\omega_k\}_{k=1}^{K}$. 

Although several individual modules are affine in their parameters, the assembled admittance is generally a nonlinear rational function of
$\theta_m$ because the modules interact through feedback
interconnections and parameter-dependent transfer functions. The
resulting least-squares problem is therefore generally nonconvex and may
contain multiple local minima. Reliable fitting is important because a
poor local solution can cause a plausible structure to receive a large error.

We use three measures to reduce this sensitivity. First, each parameter
is optimized within the admissible range associated with its module
type, as encoded by $\Theta(G)$. Second, each fit is initialized using a
vector-fitting seed obtained from the measured response and is repeated
with five additional restarts in this work. Third, when the fit at one
operating point achieves an error below a prescribed threshold, the
resulting parameter estimate is used to initialize the optimization at
the next operating point as a warm start.

\subsubsection{Reward}

After solving \eqref{eq:fit} at all $M$ operating points, the evaluator
uses the average fitting error in the outer objective
\eqref{eq:problem-out} to define the scalar reward
\begin{equation}
\label{eq:reward}
    \mathcal{R}(G)
    =
    -\log\!\left(
        \frac{1}{M}
        \sum_{m=1}^{M}
        \NRMSE_m\!\left(G,\theta_m^\star(G)\right)
        +10^{-8}
    \right).
\end{equation}
Thus, structures with lower average NRMSE receive higher reward. The
logarithmic transformation increases resolution among candidates in the
low-error regime, where differences in raw NRMSE become small.

Invalid, non-executing, or timed-out candidates receive fixed negative
rewards below those assigned to valid candidates. Structural complexity
is not included directly in \eqref{eq:reward}; it is controlled through
the hard model-order constraint \eqref{eq:problem-n}. This avoids
introducing an additional scalar tradeoff between fitting accuracy and
model complexity.

\subsubsection{Structured Feedback}

The scalar reward indicates the overall quality of a candidate but
provides limited information about the source of its modeling error.
The evaluator therefore also returns a structured residual record,
denoted by $\rho(G)$. The record resolves the fitting error along three
axes: frequency band, admittance entry, and operating point. Specifically,
the frequency-domain error is reported over
$5$--$50$\,Hz, $50$--$500$\,Hz, $0.5$--$2$\,kHz, and
$2$--$5$\,kHz; entry-wise errors are reported for
$Y_{dd}$, $Y_{dq}$, $Y_{qd}$, and $Y_{qq}$; and separate errors are
reported for each operating point $r_m$.

The record also reports the number of instantiated modules, the dynamic
order $n(G)$, the number of residual modules, and the number of free
parameters. Together with the resolved fitting errors, these quantities
provide structured feedback for subsequent proposals.

\section{Recursive Search and Self-Improvement}
\label{sec:rsi_search}

With the structured model space and evaluator of
Section~\ref{sec:model_search} fixed, the remaining task is to solve the
outer structural search in \eqref{eq:problem-out}. We address this problem
using an RSI LLM agent based on the evolutionary program-search framework
of {ThetaEvolve} \cite{wang2025thetaevolve}. Each candidate
structure $G$ is encoded as a program that instantiates modules from
$\mathcal{M}$ and connects their ports. 

Recursive improvement occurs through two coupled channels. Evaluated
structures and their evaluation records accumulate in the search archive,
changing the information available to subsequent proposals. When the
proposal model has accessible weights, evaluator rewards can additionally
update the proposal policy during search. We refer to these mechanisms as
recursion in \emph{context space} and recursion in \emph{weight space},
respectively.

\subsection{Search State and Evolutionary Archive}
\label{sub:search_state_archive}

At iteration $t$, the search state is
\begin{equation}
\label{eq:rsi_state}
    s_t=(\mathcal{A}_t,\psi_t),
\end{equation}
where $\mathcal{A}_t$ is the archive of previously evaluated candidate
structures, their rewards, and their evaluator records, and $\psi_t$
denotes the weights of the proposal policy $\pi_{\psi_t}$.

The archive maintains both high-performing candidates and
diversity. In this work, it contains a capped
population of $P=10{,}000$ candidates divided among $I=10$ islands
arranged in a ring, together with a global elite set containing the
$200$ highest-reward structures found so far. Search advances through the
islands in round-robin order. For each proposal, the parent is sampled
with probability $0.6$ from the elite set, $0.2$ from the active island,
and $0.2$ from the full population. 

After approximately $1{,}000$ admitted structures, corresponding to one
complete sweep of the $10$-island ring, each island shares its top $10\%$
of candidates with its two neighboring islands
\cite{novikov2025alphaevolve}. Once the population cap is reached,
low-reward candidates are evicted. The archive therefore balances
exploitation of high-performing structures with exploration of
structurally different alternatives.


\subsection{Proposal Context and Control-Engineering Priors}
\label{sub:proposal_context}

At iteration $t$, a parent structure $G_t^{\rm par}$ is sampled from
$\mathcal{A}_t$ according to the mechanism in
Section~\ref{sub:search_state_archive}. The proposal policy receives a
context $c_t$ containing four components: the identification objective
and evaluation metric; the graph-construction API, module library
$\mathcal{M}$, and interconnection grammar; the parent program and its
structured residual record; and control-engineering priors that guide
interpretation of the residual records.

A central prior is the relation between frequency scale and control
layer. Grid-following inverters typically contain synchronization, outer
power regulation, and inner current-control loops with distinct
bandwidths. Residuals concentrated at low frequencies can therefore
indicate missing or inaccurate synchronization or outer-loop dynamics,
while high-frequency residuals more often point to the inner current
loop or physical filter. These cues help translate the frequency-resolved
evaluator output into plausible structural modifications.

The control-engineering priors are specified independently of
measurements from the target device. They encode generic knowledge
available before identification, including standard GFL control
components, typical control-loop hierarchy, and the relative ordering of
loop bandwidths.

To balance local refinement with broader structural exploration, the search varies the scope of the requested edit. At each iteration, one of three edit scales is selected: a local edit that modifies parameter sharing or signal routing, a module-level edit that substitutes a single control component, or a larger structural edit that replaces a plant component or an entire control sub-loop. Their sampling probabilities are $0.3$, $0.4$,
and $0.3$, respectively. 

In this paper, $c_t$ contains only the sampled parent and
its evaluator record rather than several candidate programs
simultaneously. 
This design matches the context capacity of the open-weight proposal model used in
our experiments. Models with larger context windows can instead expose a
longer search history directly to the proposer.

A shortened excerpt of the proposal context used in our experiments is
shown below.

\begin{tcolorbox}[
    colback=gray!6,
    colframe=gray!40,
    boxrule=0.4pt,
    arc=1.5mm,
    left=1.5mm,
    right=1.5mm,
    top=1.3mm,
    bottom=1.3mm,
    title={\scriptsize\textbf{Context excerpt}},
    fonttitle=\scriptsize
]
\scriptsize
\textbf{Task.}
Evolve \texttt{build\_gfl\_skeleton()} to recover an interpretable
small-signal GFL inverter model from measured PCC admittance
$Y(j\omega)$.

\medskip
\textbf{Search space.}
Construct the model using only the provided typed control modules and
graph API. All connections must satisfy the interconnection grammar;
module parameters remain symbolic during structure generation and are
fitted by the evaluator.

\medskip
\textbf{Evaluation.}
The evaluator fits the continuous parameters at each operating point and
returns the NRMSE averaged across operating points, together with errors
resolved by frequency band, admittance entry, and operating point.
Lower average NRMSE corresponds to higher reward.

\medskip
\textbf{Domain cues.}
A GFL inverter typically contains synchronization, outer power control,
and inner current control loops with separated bandwidths. Use the
frequency distribution of the residual to identify which control layer
may be missing or incorrectly represented before proposing an edit.
If adding an expected control layer degrades the fit, examine its
feedback sign, measurement filtering, and injection point before
removing it.
\end{tcolorbox}

\subsection{Recursive Search Procedure}
\label{sub:recursive_search}

At iteration $t$, a parent structure $G_t^{\rm par}$ is sampled from the
archive and used to construct the proposal context $c_t$ as described in
Sections~\ref{sub:search_state_archive} and \ref{sub:proposal_context}.
The proposal policy then generates a structural edit
\begin{equation}
\label{eq:proposal_policy}
    d_t
    \sim
    \pi_{\psi_t}(\cdot\mid c_t),
\end{equation}
which is applied to the parent program to produce a candidate structure
$G_t$. In this work, proposals are represented as
localized program edits rather than regenerating the entire program.

The candidate is first checked against the module vocabulary and
interconnection grammar. Invalid candidates receive a fixed negative
reward without parameter fitting. Each valid candidate $G_t$ is evaluated according to
Section~\ref{sub:eval}, yielding the scalar reward
$\mathcal{R}_t=\mathcal{R}(G_t)$ and the structured evaluator record
$\rho_t=\rho(G_t)$. These outputs update the two components of the search state according to
\begin{equation}
\label{eq:rsi_recursion}
\begin{aligned}
    \mathcal{A}_{t+1}
    &=
    \mathcal{H}\!\left(
        \mathcal{A}_t,
        G_t,
        \mathcal{R}_t,
        \rho_t
    \right),
    \\
    \psi_{t+1}
    &=
    \mathcal{U}\!\left(
        \psi_t,
        \mathcal{D}_t
    \right),
\end{aligned}
\end{equation}
where $\mathcal{H}$ denotes the archive update defined by the admission,
eviction, and migration rules of Section~\ref{sub:search_state_archive},
and $\mathcal{D}_t$ denotes the proposal and reward data accumulated for
policy adaptation. The operator $\mathcal{U}$ is either a
reinforcement-learning update or the identity map, depending on whether
the proposal model is adapted during search. Equation~\eqref{eq:rsi_recursion}
therefore captures the two forms of recursive improvement considered in
this work: the search experience can alter the proposal policy itself,
the information supplied to future proposals, or both.

\subsection{Recursion in Weight Space: Test-Time Reinforcement Learning}
\label{sub:weight_space}

When the proposal model has accessible weights, evaluator rewards are
used to update $\pi_{\psi_t}$ during the identification process. The
resulting weight-space recursion changes the distribution of future
structural edits, allowing the proposer to learn from the outcomes of
previous candidates rather than relying on a fixed proposal policy.

We use the group-relative policy-gradient (GRPO) mechanism of
{ThetaEvolve} \cite{shao2024deepseekmath,wang2025thetaevolve} for weights update.
For each sampled context $c_t$, the policy generates a group of
$N_{\rm p}$ independent proposals with rewards
$\mathcal{R}_t^{(1)},\ldots,\mathcal{R}_t^{(N_{\rm p})}$. The advantage of
each proposal is normalized relative to the other proposals generated
from the same context:
\begin{equation}
\label{eq:grpo_advantage}
\begin{aligned}
    \widehat{A}^{(i)}
    &=
    \frac{
        \mathcal{R}_t^{(i)}-\mu_c
    }{
        \sigma_c+\epsilon
    },
    \\
    \mu_c
    &=
    \frac{1}{N_{\rm p}}
    \sum_{j=1}^{N_{\rm p}}
    \mathcal{R}_t^{(j)}.
\end{aligned}
\end{equation}
Here, $\mu_c$ and $\sigma_c$ are the mean and standard deviation of the
proposal rewards within the group, and $\epsilon=10^{-6}$ is a small
constant for numerical stability.

Let $a_1^{(i)},\ldots,a_{L_i}^{(i)}$ denote the token sequence generated
for proposal $G_t^{(i)}$, and let $\pi_{\psi_{\rm old}}$ denote the
policy used to generate the proposal batch. For each token, we define
the policy ratio
\begin{equation}
\label{eq:policy_ratio}
r_{i,\ell}(\psi)
=
\frac{
\pi_{\psi}\!\left(
a_{\ell}^{(i)} \mid c,a_{<\ell}^{(i)}
\right)
}{
\pi_{\psi_{\rm old}}\!\left(
a_{\ell}^{(i)} \mid c,a_{<\ell}^{(i)}
\right)
}.
\end{equation}

Proposal generation and policy optimization use separate serving and
training backends, which may assign slightly different probabilities to
the same generated token. We account for this mismatch using the
truncated importance weight
\begin{equation}
\label{eq:tis_weight}
w_{i,\ell}
=
\min\!\left\{
\frac{
\pi_{\rm train}\!\left(
a_{\ell}^{(i)}
\mid c,a_{<\ell}^{(i)}
\right)
}{
\pi_{\rm serve}\!\left(
a_{\ell}^{(i)}
\mid c,a_{<\ell}^{(i)}
\right)
},
\,w_{\max}
\right\}.
\end{equation}

The policy is then updated using the clipped surrogate objective
\begin{equation}
\label{eq:grpo_objective}
\begin{aligned}
\mathcal{J}_{\rm GRPO}(\psi)
&=
\frac{1}{N_{\rm p}}
\sum_{i=1}^{N_{\rm p}}
\frac{1}{L_i}
\sum_{\ell=1}^{L_i}
w_{i,\ell}
\min\Bigg(
    r_{i,\ell}(\psi)\widehat{A}^{(i)},
\\
&\qquad
    \operatorname{clip}\!\left(
        r_{i,\ell}(\psi),
        1-\epsilon_{\rm pg},
        1+\epsilon_{\rm pg}
    \right)
    \widehat{A}^{(i)}
\Bigg).
\end{aligned}
\end{equation}
The clipping parameter $\epsilon_{\rm pg}$ limits the change in token
probability during each policy update, while $w_{\max}$ limits the
influence of large probability discrepancies between the serving and
training backends.

Because proposals within a group share the same parent and proposal
context, the update directly compares alternative modifications of the
same candidate structure. Evaluator penalties discourage invalid or
ineffective edits, while higher-reward structural changes become more
likely under subsequent versions of the proposal policy.

\subsection{Recursion in Context Space: Training-Free Search}
\label{sub:context_space}

Recursive improvement can also occur without changing the proposal-model
weights. Setting $\mathcal{U}=\mathrm{id}$ in
\eqref{eq:rsi_recursion} leaves $\psi_t$ fixed, while the archive
$\mathcal{A}_t$ continues to evolve as new candidates are evaluated.
Consequently, parent selection changes over time, and future proposal
contexts contain structures and evaluator records produced by earlier
search iterations.

In the implementation considered here, this recursion is
archive-mediated: each proposal receives one sampled parent together
with its structured evaluator record, while population-level search
history influences the proposal indirectly through parent selection.
This provides a training-free search mode that can also be used with
off-the-shelf API models whose weights are unavailable.

A richer form of context-space recursion can place a capable coding
agent directly in a closed loop with the evaluator. Compared with the
proposal model used in our current implementation, such agents typically
provide larger context windows, stronger tool-use capabilities, and
broader exposure to domain knowledge. They can therefore retain a longer
history of previous hypotheses and reason over this
history when diagnosing residual errors. Within each interaction, the agent can
modify a candidate program, execute the evaluator, inspect the resulting
diagnostics, repair coding or implementation errors, and reflect before generating the next proposal. This loop closely follows the
workflow of recent autonomous-research systems
\cite{karpathy2026autoresearch,hogan2026alphalab,ishibashi2026harness}
and provides a natural future extension.

\subsection{Experimental Search Configuration}
\label{sub:search_configuration}

The reported experiments use 
DeepSeek-R1-0528-Qwen3-8B
as the proposal policy. Both recursion channels are active: the evolutionary
archive determines the parents encountered by the
proposer, while test-time reinforcement learning updates the proposal
policy from evaluator rewards.

For each policy update, $32$ proposal contexts are sampled and
$N_{\rm p}=16$ proposals are generated per context, yielding $512$
evaluated candidates. Each batch is followed by one policy update
with learning rate $10^{-6}$. An edit that leaves the parent program
unchanged is assigned a fixed reward of $-0.3$ to prevent exact parent
reproduction from receiving the parent's evaluation score.

The admissible dynamic-order range in \eqref{eq:problem-n} is
\begin{equation}
\label{eq:search_order_range}
    [\underline{n},\overline{n}]
    =
    [0,30].
\end{equation}
The search continues until the prescribed budget is exhausted.
\section{Results}
\label{sec:results}

\subsection{Setup}
\begin{figure}[!t]
    \centering
    \includegraphics[width=0.6\columnwidth]{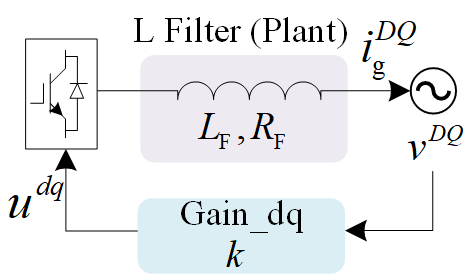}
    \caption{Na\"ive initialization of the evolutionary search. The
    initial model contains only a static gain and an L-filter plant,
    without synchronization or closed-loop inverter control.}
    \label{fig:naive_result}
\end{figure}

We evaluate the proposed RSI algorithm on the GFL inverter model of
Section~\ref{sec:prelim}. The hidden ground truth contains an LCL filter,
SRF-PLL, outer power control, inner current control, frame
transformations, and cross-coupling compensation. The measured labels
$\widehat{Y}(j\omega_k)$ are generated by frequency sweeps of the
underlying DAE over $5\,\mathrm{Hz}$--$5\,\mathrm{kHz}$ at
$p^*=1$ p.u. and two reactive-power operating points,
$q^*\in\{+0.3,-0.3\}$ p.u. The search is deliberately initialized from the simple model in
Fig.~\ref{fig:naive_result}: a static $dq$ gain feeding an L-filter
plant, with the frequency-deviation input fixed at zero. This
initialization contains none of the synchronization, outer-power, or
inner current-control loops of the hidden inverter and yields an NRMSE
of $0.470$.

\subsection{Recovered Control Structure}

\begin{figure}[!t]
    \centering
  \includegraphics[width=1\linewidth]{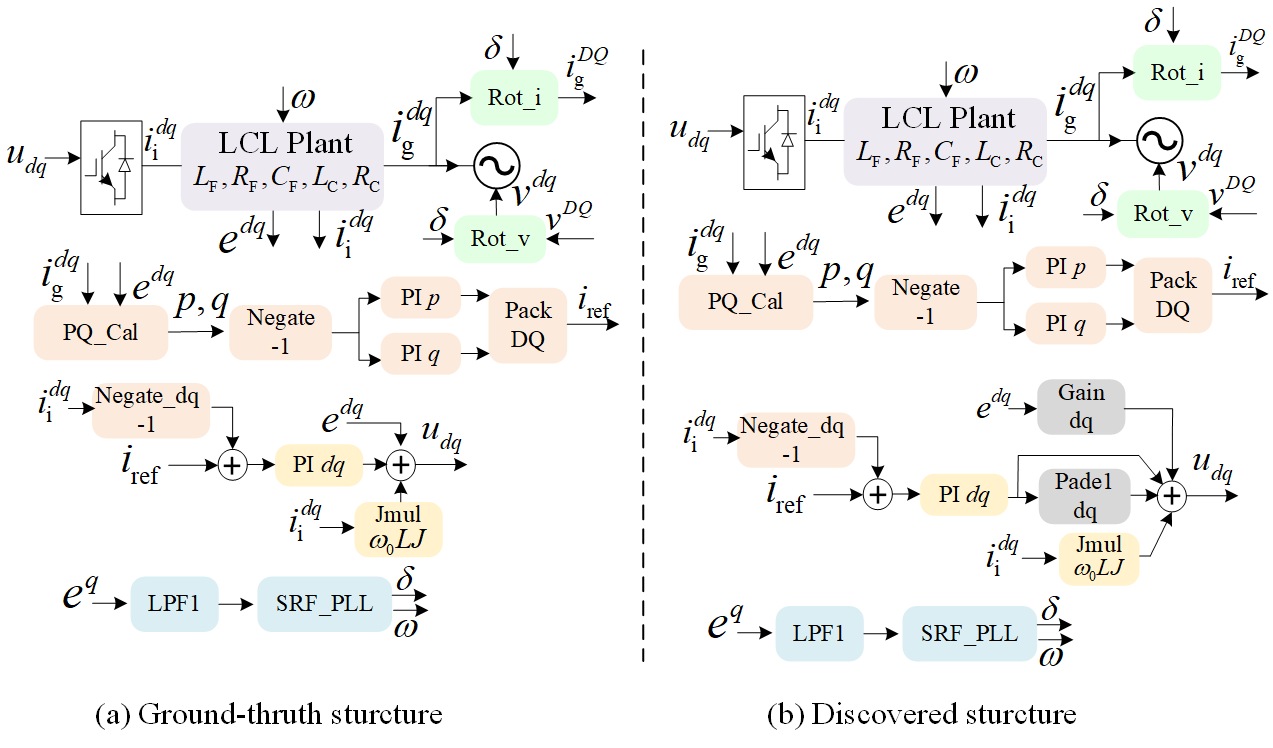}

    \caption{Comparison of the hidden ground-truth structure
    (a) and the best structure discovered by the proposed
    search (b). The recovered model reproduces the main
    synchronization, outer-power, inner-current, frame-transformation,
    and cross-coupling components. Its principal discrepancy is an
    additional gain--Pad\'e feedforward branch and a gain branch for the voltage feedforward in the inner loop (the gray box).}
    \label{fig:structure_result}
\end{figure}

Despite the deliberately weak initialization, the RSI algorithm reconstructs
nearly all major components of the hidden GFL architecture.
Figure~\ref{fig:structure_result} compares the ground-truth and recovered
block diagrams. Both contain Park transformations at the PCC interfaces,
an LCL plant, an LPF-conditioned PLL synchronization loop, PQ calculation
with separate active- and reactive-power PI controllers, a packed $dq$
inner current controller, and $\Jmat$-based cross-coupling compensation.
The principal structural discrepancy is localized to the inner-loop
feedforward path: the recovered model introduces an additional gain and
Pad\'e-delay branch that is absent from the ground truth, as shown in the gray box in Fig. \ref{fig:structure_result} (b).

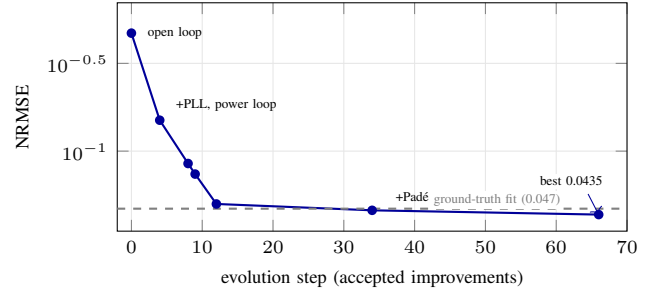
\begin{figure}[!t]
\centering
\begin{tikzpicture}
\begin{axis}[
  width=0.94\columnwidth,
  height=4.6cm,
  xlabel={evolution step (accepted improvements)},
  ylabel={NRMSE},
  ymode=log,
  xmin=-2,
  xmax=70,
  ymin=0.035,
  ymax=0.7,
  grid=both,
  grid style={gray!20},
  tick label style={font=\scriptsize},
  label style={font=\scriptsize},
  every axis plot/.append style={thick},
]

\addplot[
  color=blue!60!black,
  mark=*,
  mark size=1.4pt
] coordinates {
  (0,0.470)
  (4,0.150)
  (8,0.085)
  (9,0.074)
  (12,0.050)
  (34,0.046)
  (66,0.0435)
};

\node[
  font=\tiny,
  anchor=west
] at (axis cs:1,0.470)
{open loop};

\node[
  font=\tiny,
  anchor=south west
] at (axis cs:5,0.150)
{+PLL, power loop};

\node[
  font=\tiny,
  anchor=south west
] at (axis cs:36,0.046)
{+Pad\'e};

\addplot[
  dashed,
  gray
] coordinates {
  (-2,0.047)
  (70,0.047)
};

\node[
  font=\tiny,
  gray,
  anchor=south east,
  fill=white,
  inner sep=1pt
] at (axis cs:61,0.047)
{ground-truth fit (0.047)};

\node[
  font=\tiny,
  anchor=south east,
  fill=white,
  inner sep=1pt
] (bestlabel) at (axis cs:67,0.061)
{best 0.0435};

\draw[
  ->,
  thin,
  color=blue!60!black
]
(axis cs:64.5,0.057)
-- (axis cs:66,0.0435);

\end{axis}
\end{tikzpicture}

\caption{Evolution trajectory of the best program. Each accepted
structural edit removes a mismatch in a predictable frequency band; the
final model slightly outperforms a parameter fit of the ground-truth
structure itself.}
\label{fig:trajectory}
\end{figure}

The emergence of this structure is accompanied by a substantial
reduction in frequency-domain error. Table~\ref{tab:trajectory}
summarizes representative accepted structures along the search
trajectory, while Fig.~\ref{fig:trajectory} shows the evolution of the
best NRMSE. Starting from the na\"ive open-loop model with NRMSE
$0.470$, the largest early improvement occurs when synchronization and
outer power-control structure is introduced. Subsequent edits recover
additional frame-handling, current-control, and high-frequency dynamics,
progressively reducing the error into the sub-$0.05$ regime.

The best recovered model contains 15 modules with declared dynamic
order 15 and achieves an NRMSE of $0.0435$. For reference, nonlinear
least-squares fitting of the known ground-truth block structure yields
an NRMSE of $0.047$. Thus, the final model achieves essentially the same
terminal behavior as the ground-truth architecture while being recovered
from PCC admittance measurements alone.


The slightly lower error of the discovered model should not be interpreted as evidence of more accurate structural recovery. As Fig.~\ref{fig:structure_result} shows, the discovered structure contains an additional gain--Pad\'e branch that is absent from the true
controller. Because the model parameters are fitted independently at each operating point, this additional flexibility can absorb residual
modeling mismatch and reduce the frequency-domain error without corresponding to the true physical structure. The result therefore highlights a limitation of the current RSI evaluation setup: per-operating-point fitting can favor more flexible structures, and excellent agreement with the measured admittance does not by itself guarantee structural uniqueness.

\begin{table}[!t]
\caption{Milestones Along the Evolution Trajectory}
\label{tab:trajectory}
\centering
\scriptsize
\setlength{\tabcolsep}{3.5pt}
\begin{tabular}{@{}rlcl@{}}
\toprule
Step & Module(s) added & NRMSE & Layer / band fixed \\
\midrule
0  & LCL plant $+$ open-loop gain & 0.470 & physical filter only \\
4  & SRF-PLL $+$ PQ calc $+$ power PI & 0.150 & sync $+$ outer loop (5--500\,Hz) \\
8  & LPF ($+$ JMul decoupling) & 0.085 & meas.\ filter, decoupling \\
9  & Park rotations & 0.074 & frame linearization \\
12 & inner-loop parameter refit & 0.050 & target band ($<0.05$) reached \\
34 & Pad\'e delay & 0.046 & delay residual (2--5\,kHz) \\
66 & JMul (final decoupling) & \textbf{0.0435} & best; order 15, 15 modules \\
\bottomrule
\end{tabular}
\end{table}

\subsection{Frequency-Domain Validation}

\begin{figure}[!t]
    \centering
    \includegraphics[width=\columnwidth]{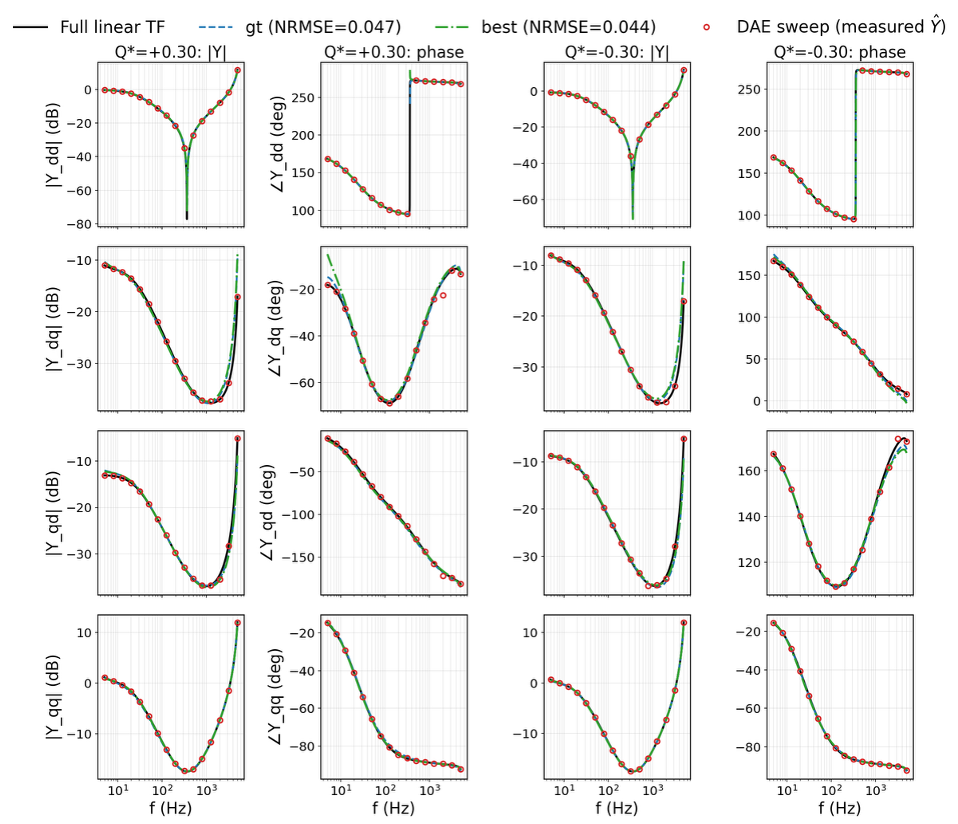}
    \caption{Frequency-domain validation at
    $q^*=+0.3$ and $q^*=-0.3$ p.u. Red markers denote the measured DAE
    sweeps, the black curves the full linearized model, the dashed blue
    curves the fitted ground-truth structure, and the dash-dotted green
    curves the best discovered structure. The recovered model closely
    matches the measured magnitude and phase across both operating
    points and all admittance channels.}
    \label{fig:bode_result}
\end{figure}

Figure~\ref{fig:bode_result} compares the frequency responses of the
recovered model with the measured DAE sweeps and the fitted
ground-truth structure. Across both operating points and all four
entries of the $2\times2$ admittance matrix, the recovered model closely
tracks the measured magnitude and phase over the full frequency range.

This agreement is significant because the search observes only the PCC
admittance and never receives the hidden control interconnections.
Together with Fig.~\ref{fig:structure_result}, the result shows that the
search recovers both a physically recognizable controller architecture
and its corresponding terminal dynamics.

\section{Conclusion and Discussion}
\label{sec:conclusion}

This paper argues that LLM-driven evolutionary search over a typed
vocabulary of control modules provides a promising framework for
grey-box identification of black-box inverters. In our proof-of-concept
GFL benchmark, a {ThetaEvolve}-based search starts from a
deliberately na\"ive model and recovers the main synchronization,
outer-power, and inner-current control structure from PCC admittance
measurements alone. The resulting model reduces NRMSE from $0.470$ to
$0.0435$ and closely reproduces the measured frequency response across
both operating points.

\subsection{Limitations}



The present results are preliminary. We consider a single simulated GFL
inverter whose underlying components are representable by the module
vocabulary, use clean frequency-domain measurements, and evaluate only
two operating points. Robustness to measurement noise, model mismatch,
unseen inverter architectures, and hardware measurements remains to be
established.

The current implementation also fits the continuous parameters
independently at each operating point. The experiment therefore
primarily evaluates recovery of a shared control structure rather than
identification of a single globally parameterized inverter model. This
per-operating-point fitting can also favor more flexible structures. In
the reported example, the discovered model contains an additional
gain--Pad\'e branch that is absent from the true controller yet slightly
reduces the fitting error. The additional flexibility allows the model
to absorb residual mismatch separately at each operating point without
corresponding to the true physical structure.

More fundamentally, terminal measurements may not uniquely determine the internal controller. A finite set of PCC frequency responses can
leave some internal dynamics weakly observable or unidentifiable, and distinct control structures may produce nearly indistinguishable
admittances. The current search also does not impose properties such as passivity, stability margins, or robustness constraints, which provide additional directions for constraining the identified models.

\subsection{RSI Beyond the Present Configuration}
The two recursion channels suggest natural extensions of the present setup. In context space, a stronger variant could place a large coding agent in a closed loop with the evaluator as described in Section \ref{sub:context_space}.
For industrial deployment, however, API-based agents introduce trade-offs in data locality, reproducibility, and inference cost.

In weight space, {ThetaEvolve} enables test-time reinforcement
learning from evaluator feedback \cite{wang2025thetaevolve}, and this
mechanism is active in our experiments. We do not isolate its contribution
from archive-based context recursion here. A controlled comparison between
frozen-policy search and test-time adaptation is therefore an important
next step, particularly for search efficiency and structural recovery.
The two channels are complementary:
context recursion can operate alone or alongside weight updates, making
the same framework applicable to both open-weight and API-based models.

\subsection{Outlook}

Several extensions are especially important. Future identification should share physical and controller parameters across operating points while allowing equilibrium-dependent quantities to vary, and should use additional operating conditions and held-out measurements to strengthen tests of structural and parametric identifiability.

The search space should also be extended beyond a single GFL family to GFM controllers, hybrid architectures, and multiple devices behind a common PCC \cite{wang2026unified,askarian2024multimode}. Selected nonlinear modules, such as saturation, current limiting, and protection logic, could further extend structural discovery beyond the small-signal regime.

Finally, the evaluator should incorporate criteria beyond frequency-response fit. The recovered modular architecture could then serve as a scaffold for state-space or nonlinear models suitable for simulation and stability analysis. Establishing when such models are identifiable, transferable, and reliable on field data is the key step toward practical, measurement-driven discovery of interpretable
inverter models.

\bibliographystyle{IEEEtran}
\bibliography{IEEEabrv,mybibfile}

\appendices
\section{Module Vocabulary}
\label{app:vocab}

This appendix specifies every module kind in the vocabulary $\mathcal{M}$ used
in Table~\ref{tab:vocab}. All quantities are small-signal perturbations in per
unit, $s=j\omega$, and $\omega_0=2\pi\cdot 60$~rad/s is the nominal angular
frequency. Throughout,
\begin{equation}
    I_2=\begin{bmatrix}1&0\\0&1\end{bmatrix},\qquad
    J=\begin{bmatrix}0&-1\\1&0\end{bmatrix}.
\end{equation}

\subsection{Signals, ports, and composition}
\label{app:vocab-composition}

Every port carries one of three signal kinds, and a connection is admissible
only between two ports of the same kind:
\begin{center}
\begin{tabular}{@{}lll@{}}
\toprule
Kind & Meaning & Width \\
\midrule
\texttt{scalar} & one signal & $1$ \\
\texttt{dq\_vec} & direct- and quadrature-axis pair & $2$ \\
\texttt{power\_pair} & active- and reactive-power pair & $2$ \\
\bottomrule
\end{tabular}
\end{center}

A module of kind $\kappa$ with parameters $p$ contributes the linear relation
\begin{equation}
\label{eq:module-relation}
    y_\kappa = M_\kappa(p,s)\,u_\kappa,
\end{equation}
where $u_\kappa$ and $y_\kappa$ stack its input and output ports in the order
listed below. Each connection imposes the constraint that the destination port
equals the sum of its source ports, which is how summing junctions and negative
feedback are expressed: subtraction is a \texttt{Negate} in series with a sum.
At each frequency the module relations \eqref{eq:module-relation} and the
connection constraints form one complex linear system; exciting the external
port voltage along the direct and then the quadrature axis and reading the
external grid current gives the $2\times2$ admittance $Y_G(j\omega;\theta)$.

Two structural quantities are attached to each kind. Its \emph{order} is the
number of dynamic states it declares, and the sum of these over a candidate is
the model order $n(G)$ bounded in \eqref{eq:problem}. Its \emph{parameter box}
is the admissible range of each parameter. Parameters are referred to by name,
so a candidate may bind the same fitted scalar to several modules; when it
does, the effective box is the intersection of the individual boxes. This is
how the steady-state quantities $E_{0d}$, $I_{0d}$, and $\delta_{\mathrm{eq}}$
are kept consistent between the plant, the rotations, and the power
calculation.

\subsection{Physical plant}
\label{app:vocab-plant}

\subsubsection{\texttt{LCL} (order 6)}
The inductor--capacitor--inductor output filter in the controller frame, with
states $x=[\,i_{ti}^{dq},\,e^{dq},\,i_g^{dq}\,]^\top$: converter-side current,
capacitor voltage, and grid-side current. Inputs are the converter voltage
command \texttt{u\_r} (\texttt{dq\_vec}), the terminal voltage \texttt{v\_in}
(\texttt{dq\_vec}), and the frame frequency perturbation \texttt{delta\_omega}
(\texttt{scalar}). Outputs are \texttt{i\_ti}, \texttt{e}, \texttt{i\_g}
(\texttt{dq\_vec}), \texttt{e\_q} (\texttt{scalar}, the quadrature component of
the capacitor voltage), and \texttt{i\_C} (\texttt{dq\_vec}, the capacitor
current $i_{ti}-i_g$). Its response is $M_{\mathrm{LCL}}=C(sI_6-A)^{-1}B$ with
\begin{equation}
    A=\begin{bmatrix}
    -\tfrac{\omega_0 R_F}{L_F}I_2-\omega_0 J & -\tfrac{\omega_0}{L_F}I_2 & 0\\[2pt]
    \tfrac{\omega_0}{C_F}I_2 & -\omega_0 J & -\tfrac{\omega_0}{C_F}I_2\\[2pt]
    0 & \tfrac{\omega_0}{L_\tg}I_2 & -\tfrac{\omega_0 R_\tg}{L_\tg}I_2-\omega_0 J
    \end{bmatrix},
\end{equation}
\begin{equation}
    B=\begin{bmatrix}
    \tfrac{\omega_0}{L_F}I_2 & 0 & -J\,i_{ti}^{\mathrm{eq}}\\[2pt]
    0 & 0 & -J\,e^{\mathrm{eq}}\\[2pt]
    0 & -\tfrac{\omega_0}{L_\tg}I_2 & -J\,i_g^{\mathrm{eq}}
    \end{bmatrix}.
\end{equation}
The third column of $B$ is the frame-rotation coupling: the controller frame
turns at $\omega_0+\Delta\omega$, and linearizing the term $-\omega Jx$ leaves
$-Jx^{\mathrm{eq}}\Delta\omega$. Under the unity-power-factor approximation the
equilibrium vectors are taken as $i_{ti}^{\mathrm{eq}}=i_g^{\mathrm{eq}}=
[I_{0d},0]^\top$ and $e^{\mathrm{eq}}=[E_{0d},0]^\top$, so each such column is
$[0,-I_{0d}]^\top$ or $[0,-E_{0d}]^\top$. Parameters and ranges:
$L_F,L_C\in[10^{-4},1]$, $R_F,R_C\in[10^{-5},1]$, $C_F\in[10^{-5},1]$,
$E_{0d},I_{0d}\in[0.5,1.5]$.

\subsubsection{\texttt{L Filter} (order 2)}
The reduced inductor-only filter, with state $i_g^{dq}$ and the same three
inputs. Outputs are \texttt{i\_g} (\texttt{dq\_vec}), \texttt{e}
(\texttt{dq\_vec}), and \texttt{e\_q} (\texttt{scalar}); since there is no
capacitor, \texttt{e} and \texttt{e\_q} are the terminal voltage \texttt{v\_in}
and its quadrature component fed straight through. With
\begin{equation}
    A=-\tfrac{\omega_0 R_F}{L_F}I_2-\omega_0 J,\qquad
    B=\begin{bmatrix}\tfrac{\omega_0}{L_F}I_2 & -\tfrac{\omega_0}{L_F}I_2 &
    -J\,[I_{0d},0]^\top\end{bmatrix},
\end{equation}
the current response is $(sI_2-A)^{-1}B$ and the voltage outputs are the
corresponding feedthrough. Parameters: $L_F\in[10^{-4},1]$,
$R_F\in[10^{-5},1]$, $I_{0d}\in[0.5,1.5]$.

Every candidate must contain exactly one plant module, either \texttt{LCL} or
\texttt{LFilter}; this is enforced by the structure validator.

\subsection{Dynamic control}
\label{app:vocab-control}

These are the textbook single-input single-output blocks and their
axis-decoupled two-axis counterparts. A two-axis block applies the same scalar
response independently on both axes, that is $M=g(s)I_2$, and declares twice
the order for that reason.

\begin{center}
\begin{tabular}{@{}llcl@{}}
\toprule
Kind & Ports & Order & Response \\
\midrule
\texttt{PI} & scalar $\to$ scalar & $1$ & $k_P+k_I/s$ \\
\texttt{PI\_dq} & dq $\to$ dq & $2$ & $(k_P+k_I/s)I_2$ \\
\texttt{LPF1} & scalar $\to$ scalar & $1$ & $\omega_c/(s+\omega_c)$ \\
\texttt{LPF1\_dq} & dq $\to$ dq & $2$ & $\bigl(\omega_c/(s+\omega_c)\bigr)I_2$ \\
\texttt{Integrator} & scalar $\to$ scalar & $1$ & $1/s$ \\
\bottomrule
\end{tabular}
\end{center}

Ranges are $k_P\in[10^{-6},10^{5}]$, $k_I\in[10^{-6},10^{7}]$, and
$\omega_c\in[2\pi\cdot 0.1,\;2\pi\cdot 5\times10^{4}]$~rad/s;
\texttt{Integrator} has no parameters. The input port is named \texttt{in\_}
and the output port \texttt{out} in all five cases.

\subsection{Synchronization}
\label{app:vocab-sync}

\subsubsection{\texttt{SRF\_PLL} (order 2)}
The type-2 synchronous-reference-frame phase-locked loop, consisting of a
proportional-integral stage followed by an integrator. Its input is the
quadrature voltage error \texttt{e\_q} (\texttt{scalar}) and it has two scalar
outputs, the frequency perturbation \texttt{delta\_omega} and the angle
perturbation \texttt{delta}:
\begin{equation}
    \begin{bmatrix}\Delta\omega\\ \delta\end{bmatrix}
    =\begin{bmatrix}
    k_P^{\mathrm{pll}}+k_I^{\mathrm{pll}}/s\\[3pt]
    \bigl(k_P^{\mathrm{pll}}+k_I^{\mathrm{pll}}/s\bigr)/s
    \end{bmatrix}\Delta e_q .
\end{equation}
Any pre-filter on the error, such as \texttt{LPF1} or \texttt{Notch}, is not
built in and must be chained externally by the candidate, which is one of the
structural choices the search can make. The gains are named
$k_P^{\mathrm{pll}}$ and $k_I^{\mathrm{pll}}$, distinct from the control-loop
gains so that a candidate cannot bind them to the same fitted value by
accident; ranges match those of \texttt{PI}.

\subsection{Operating-point modules}
\label{app:vocab-oppoint}

These carry the dependence on the steady-state operating condition. They are
memoryless, so their order is zero, but their entries are functions of the
fitted equilibrium quantities $\delta_{\mathrm{eq}}\in[-0.5,0.5]$~rad,
$E_{0d}\in[0.5,1.5]$, and $I_{0d}\in[0.5,1.5]$. The port voltage in the global
frame is $V_0^{DQ}=[1,0]^\top$ by convention, and the phase-locked loop drives
the quadrature components of the equilibrium to zero.

\subsubsection{\texttt{Rot\_v} (order 0)}
Rotation of the terminal voltage from the global frame into the controller
frame, linearized at $\delta_{\mathrm{eq}}$. Inputs \texttt{v\_DQ}
(\texttt{dq\_vec}) and \texttt{delta} (\texttt{scalar}), output \texttt{v\_dq}
(\texttt{dq\_vec}):
\begin{equation}
    M_{\texttt{Rot\_v}}=
    \begin{bmatrix}
    \cos\delta_{\mathrm{eq}} & \sin\delta_{\mathrm{eq}} & -\sin\delta_{\mathrm{eq}}\\
    -\sin\delta_{\mathrm{eq}} & \cos\delta_{\mathrm{eq}} & -\cos\delta_{\mathrm{eq}}
    \end{bmatrix}.
\end{equation}
The third column is the sensitivity of the rotation to the angle perturbation
supplied by the phase-locked loop.

\subsubsection{\texttt{Rot\_i} (order 0)}
The inverse rotation applied to the grid current. Inputs \texttt{i\_dq}
(\texttt{dq\_vec}) and \texttt{delta} (\texttt{scalar}), output \texttt{i\_DQ}
(\texttt{dq\_vec}):
\begin{equation}
    M_{\texttt{Rot\_i}}=
    \begin{bmatrix}
    \cos\delta_{\mathrm{eq}} & -\sin\delta_{\mathrm{eq}} & -I_{0d}\sin\delta_{\mathrm{eq}}\\
    \sin\delta_{\mathrm{eq}} & \cos\delta_{\mathrm{eq}} & \phantom{-}I_{0d}\cos\delta_{\mathrm{eq}}
    \end{bmatrix}.
\end{equation}

\subsubsection{\texttt{PQ\_calc} (order 0)}
Instantaneous power, linearized at the fitted equilibrium. Inputs \texttt{e}
and \texttt{i} (both \texttt{dq\_vec}, stacked as
$[\Delta e_d,\Delta e_q,\Delta i_d,\Delta i_q]^\top$), outputs \texttt{p} and
\texttt{q} (both \texttt{scalar}):
\begin{equation}
    \begin{bmatrix}\Delta p\\ \Delta q\end{bmatrix}
    =\begin{bmatrix}
    I_{0d} & 0 & E_{0d} & 0\\
    0 & I_{0d} & 0 & -E_{0d}
    \end{bmatrix}
    \begin{bmatrix}\Delta e_d\\ \Delta e_q\\ \Delta i_d\\ \Delta i_q\end{bmatrix}.
\end{equation}

\subsection{Static algebra}
\label{app:vocab-static}

Memoryless blocks that carry no dynamics; all have order $0$. They exist so
that wiring choices such as sign inversion, axis selection, and cross-axis
decoupling are explicit parts of the structure rather than hidden conventions.

\begin{center}
\begin{tabular}{@{}llll@{}}
\toprule
Kind & Ports & Response & Parameters \\
\midrule
\texttt{Gain} & scalar $\to$ scalar & $k$ & $k\in[-100,100]$ \\
\texttt{Gain\_dq} & dq $\to$ dq & $kI_2$ & $k\in[-100,100]$ \\
\texttt{Negate} & scalar $\to$ scalar & $-1$ & --- \\
\texttt{Negate\_dq} & dq $\to$ dq & $-I_2$ & --- \\
\texttt{SliceD} & dq $\to$ scalar & $[1\;\;0]$ & --- \\
\texttt{SliceQ} & dq $\to$ scalar & $[0\;\;1]$ & --- \\
\texttt{PackDQ} & 2 scalars $\to$ dq & $I_2$ & --- \\
\texttt{JMul} & dq $\to$ dq & $\omega_0 L\,J$ & $L\in[10^{-5},5]$ \\
\texttt{ConstZero} & --- $\to$ scalar & $0$ & --- \\
\bottomrule
\end{tabular}
\end{center}

\texttt{PackDQ} takes its two scalar inputs as \texttt{in\_d} and
\texttt{in\_q} and emits them as the two components of one two-axis signal;
\texttt{SliceD} and \texttt{SliceQ} are its inverses. \texttt{JMul} is the
standard cross-axis decoupling term $\omega_0 L\,J$, which maps the
direct-axis current onto the quadrature-axis voltage command and vice versa.
\texttt{ConstZero} is a source with no input, used to tie a reference or an
unused summing input to zero, which the validator requires to be driven
exactly once like any other input.

\subsection{Residual modules}
\label{app:vocab-residual}

These three do not correspond to a controller block a designer would draw.
They absorb effects such as computation and modulation delay or an unmodeled
resonance, and they are flagged separately so that a candidate relying on them
can be recognized as leaning on unstructured correction rather than on
identified control structure.

\begin{center}
\begin{tabular}{@{}llcl@{}}
\toprule
Kind & Ports & Order & Response \\
\midrule
\texttt{Pade1} & scalar $\to$ scalar & $1$ &
    $\dfrac{1-T_ds/2}{1+T_ds/2}$ \\[6pt]
\texttt{Pade1\_dq} & dq $\to$ dq & $2$ &
    $\dfrac{1-T_ds/2}{1+T_ds/2}I_2$ \\[6pt]
\texttt{Notch} & scalar $\to$ scalar & $2$ &
    $\dfrac{s^2+\omega_n^2}{s^2+(\omega_n/Q)s+\omega_n^2}$ \\
\bottomrule
\end{tabular}
\end{center}

The first two are the first-order Pad\'e approximation of a delay $T_d$, with
$T_d\in[10^{-6},5\times10^{-3}]$~s. The notch has
$\omega_n\in[2\pi\cdot 5,\;2\pi\cdot 10^{4}]$~rad/s and $Q\in[0.5,100]$.

\end{document}